\documentclass[aps,prl,floats,onecolumn,showpacs,superscriptaddress,floatfix]{revtex4}

\usepackage{graphicx,epsfig}
\usepackage{times}
\usepackage{graphics,dcolumn,bm,fleqn,epic,eepic,float}
\usepackage{amssymb,amsmath,multirow,rotate,color}
\begin{document}

\preprint{APS/123-QED}

\title{Fence-sitters Protect Cooperation in Complex Networks}

\author{Yichao Zhang}
%\email{yiczhang@cs.ucl.ac.uk}
\affiliation{Univ Normandy, France; ULH, LMAH, F-76600 Le Havre, FR CNRS 3335, ISCN, 25 rue Philippe Lebon, 76600 Le Havre, France}
\affiliation{Department of Computer Science, University
College London, Gower Street, London, WC1E 6BT, United Kingdom}

\author{M. A. Aziz-Alaoui}
\email{aziz.alaoui@univ-lehavre.fr}
\affiliation{Univ Normandy, France; ULH, LMAH, F-76600 Le Havre, FR CNRS 3335, ISCN, 25 rue Philippe Lebon, 76600 Le Havre, France}

\author{Cyrille Bertelle}
%\email{cyrille.bertelle@univ-lehavre.fr}
\affiliation{Univ Normandy, France; ULH, LITIS, ISCN, F-76600 Le Havre, 25 rue Philippe Lebon, 76600 Le Havre, France}

\author{Shi Zhou}
%\email{s.zhou@ucl.ac.uk}
\affiliation{Department of Computer Science, University
College London, Gower Street, London, WC1E 6BT, United Kingdom}

\author{Wenting Wang}
%\email{wenting.wang.10@ucl.ac.uk}
\affiliation{Department of Mathematics, University
College London, Gower Street, London, WC1E 6BT, United Kingdom}

\date{\today}
\begin{abstract}
Evolutionary game theory is one of the key paradigms behind many scientific disciplines from science to engineering. In complex networks, because of the difficulty of formulating the replicator dynamics, most of previous studies are confined to a numerical level. In this paper, we introduce a vectorial formulation to derive three classes of individuals' payoff analytically. The three classes are pure cooperators, pure defectors, and fence-sitters. Here, fence-sitters are the individuals who change their strategies at least once in the strategy evolutionary process. As a general approach, our vectorial formalization can be applied to all the two-strategies games. To clarify the function of the fence-sitters, we define a parameter, payoff memory, as the number of rounds that the individuals' payoffs are aggregated. We observe that the payoff memory can control the fence-sitters' effects and the level of cooperation efficiently. Our results indicate that the fence-sitters' role is nontrivial in the complex topologies, which protects cooperation in an indirect way. Our results may provide a better understanding of the composition of cooperators in a circumstance where the temptation to defect is larger.
\end{abstract}

\maketitle

\section{Introduction}
In the past half century, evolutionary game theory made a series of remarkable advances~\cite{ETG,GTE,BAMS40479}. It has finally matured into a widely accepted research tool in natural and social sciences. The tool effectively promoted development of these disciplines. Compared with the well-mixed unstructured populations~\cite{GTE,EGPD}, recent results indicate that the cooperators obtain a larger living space in the complex networks~\cite{PRL95098104,PRL98108103,PRE72056128,PRE66021907,EPL7730004,NATURE359826,NATURE433312,NATURE441502,SCIENCE3141560}. Previous numerical studies show the influence of topological structures on the level of cooperation. A variety of structures were then intensively investigated~\cite{PR44697}, such as regular graphs~\cite{NATURE359826,NATURE441502}, small-world networks~\cite{PRL95098104,PRE72056128}, random graphs~\cite{NATURE433312,NATURE441502}, and scale-free networks~\cite{NATURE433312,NATURE441502,PRL95098104,PRL98108103}. In these topological structures, the group of opponents surrounding a node is interpreted as its neighbors. The limited local interactions are the interactions restricted among the node and its neighbors.

In this paper, we investigate the role of fence-sitters in evolutionary games in complex topologies. Fence-sitters are the individuals who change their strategies at least once in the strategy evolutionary process. Considering the difficulty of formulating the strategy updating process in complex topologies, we provide a tool for deriving the pure cooperators, pure defectors and fence-sitters' payoffs.
To stay consistent with the previous studies, we adopt the prisoner's dilemma (PD) as the game model. As a heuristic framework, the prisoner's dilemma describes a commonly identified paradigm in many real-world situations~\cite{GTE,ETG,EC,JCR,GCEC,NATURE398441}. It has been widely studied as a standard model for the confrontation between cooperative and selfish behaviors.
The selfish behavior here is manifested by a defective strategy, aspiring to obtain the greatest benefit from the interaction with others. For a two-strategies game as the PD, an individual gains $T$ (temptation to defect) for defecting a cooperator, $R$ (reward for mutual cooperation) for cooperating with a cooperator, $P$ (punishment for mutual defection) for defecting a defector and $S$ (sucker's payoff) for cooperating with a defector. For the PD, the four payoff values are specifically defined as: $T>R>P\geqslant S$. At the next round, the individual will know the strategy of the other one in the previous round. It can then adjust its strategy according to the updating rule of strategy.
%
%This PD game model~\cite{ETG,GTE} considers two prisoner's who are placed in separate cells. Each prisoner must decide to confess (defect) or keep silence (cooperate). A prisoner may receive one of the following four different payoffs depending on both its own strategy and the other prisoner's strategy. It gains $T$ (temptation to defect) for defecting a cooperator, $R$ (reward for mutual cooperation) for cooperating with a cooperator, $P$ (punishment for mutual defection) for defecting a defector and $S$ (sucker's payoff) for cooperating with a defector. Normally, the four payoff values are defined as: $T>R>P\geqslant S$. At the next round, the prisoner will know the strategy of the other one in the previous round. It can then adjust its strategy according to the game model. In this game model there is a safe strategy, i.e., a prisoner $i$ always gets a higher payoff if he chooses to defect.

\section{Payoff of Individuals}
To derive the payoff of the pure strategists and fence-sitters, we will introduce a mean-field approach in the following. In the PD, $i$'s strategy can be denoted by $\Omega_i$. $\Omega_i$ takes vectors ${(1,0)}^\textrm{tr}$ and ${(0,1)}^\textrm{tr}$ for the cooperative and defective strategy respectively.
In a two-strategies evolutionary game, $i$'s strategy $\Omega_i(n)$ can be represented as
\begin{equation}
\left(
  \begin{array}{cc}
    X_i(n) \\
    1-X_i(n) \\
  \end{array}
\right).\label{X_n}
\end{equation}
$X_i(n)$ can only take $1$ or $0$ at the $n$th round. $i$ is a cooperator (defector) for $X_i(n)=1$ ($0$). Note that this vectorial formulation is completely different from the ``mixed strategy'' in replicator equations~\cite{BAMS40479,PRL97158701,PRL103198702,JSTAT08007}. The element in the mixed strategy denotes the probability of adopting this pure strategy, while our definition of $\Omega_i(n)$ is to facilitate the following calculations of vector. Thus $X_i(n)$ itself does not contain any physical meaning.

At the $n$th round of game playing with the other prisoner $j$, $i$'s payoff $G_i(n)$ can be rewritten as
\begin{equation}
G_i(n)={\Omega_i^\textrm{tr}}\left(
  \begin{array}{cc}
    R & S\\
    T & P\\
  \end{array}
\right){\Omega_j}.
\end{equation}
In a structured group, an individual $i$ plays with $k_i$ neighbors, where $k_i=\sum_jA_{ij}$ denotes $i$'s degree or connectivity. $A_{ij}$ is an entry of the adjacent matrix of networks, taking values $A_{ij}=1$ ($i=1$, $2$, $...$, $the~size~of~networks$) whenever node $i$ and $j$ are connected and $A_{ij}=0$ otherwise.
In the evolutionary process, the PD is repeated in the following way: in the first stage, a PD is played by every pair of nodes connected by a link in the network. For a node $i$ in the $n$th round, its payoff reads as
\begin{equation}
G_i(n)={\Omega_i^\textrm{tr}(n)}\left(
  \begin{array}{cc}
    R & S\\
    T & P\\
  \end{array}
\right)\sum_jA_{ij}{\Omega_j(n)}.\label{G_i}
\end{equation}
Each node then updates its accumulated payoff, which is the sum of payoffs it receives from the last rounds in memory. We define $\lambda$ as the span of the payoff memory. For $\lambda=5$, an individual's payoff keeps aggregating for five rounds. At the beginning of the sixth round, the whole payoff system is reset. The purpose of introducing the payoff memory into the model is to simulate a points system of aggregating the players' payoff. For example, a season in the English Premier League normally lasts $38$ rounds, in which $38$ denotes the memory span. In the second stage, each node updates its strategy based on its aggregated payoff. According to various replicator dynamics, the node will adjust its strategy in the next round. Note that, the payoff memory mentioned here is different from the memory in ``memory loss'' defined in Refs.~\cite{PRL103198702,JSTAT08007}. In these two references, the memory denotes the aggregated fitness of individuals. It is also different from the ``Time scales'' in Ref.~\cite{PRL97158701}, which denotes the time spans of selections and interactions in one round of game.

For $i$'s local gaming environment, we define $W_i(n)=\frac{\sum_jA_{ij}{\Omega_j^\textrm{tr}(n)}\left(1~0\right)^\textrm{tr}}{k_i}$
as $i$'s local frequency of cooperators at the $n_{th}$ round. For the global gaming environment, we define $Q(n)=\frac{\sum_i{\Omega_i^\textrm{tr}(n)}\left(1~0\right)^\textrm{tr}}{N}$ as the global frequency of cooperators at the $n$th round. $N$ denotes the size of network. To understand exactly the role of fence-sitters, we choose random graphs (networks) as the gaming platform where the distribution of individuals playing a strategy can be served as an arbitrary distribution for one thing. For another, the large amount of connections between the pure cooperators and fence-sitters~\cite{PRL98108103} lead to the fact that the fraction of defectors is sensitive to the fence-sitters' higher payoff. The fence-sitter denotes an individual changing its strategy at least once in the evolutionary process. The comparison among the degree distributions of cooperators, defectors and all the individuals numerically will be shown in our simulation section to clarify this point. In this case,
\begin{equation}
W_i(n)\simeq Q(n).\label{Approxiation}
\end{equation}
By this mean-field approximation, $\sum_jA_{ij}{\Omega_j(n)}$ in random graph can be rewritten as
\begin{equation}
\sum_j\!\! A_{ij}{\Omega_j(n)}\!=\!{\!\langle{k}\rangle}\!\left(\!\!\left(
  \begin{array}{cc}
    1 \\
    0 \\
  \end{array}
\right)Q(n)
+\left(
  \begin{array}{cc}
    0 \\
    1 \\
  \end{array}
\right)(1-Q(n))\!\!\right),\label{sum_j}
\end{equation}
where $\langle{k}\rangle$ denotes the average degree of all the individuals.
Inserting Eq.~(\ref{sum_j}) and Eq.~(\ref{X_n}) into Eq.~(\ref{G_i}), we have
\begin{eqnarray}
&&G_i(n)=\langle{k}\rangle\left\{\left[S-P+(R-T+P-S)Q(n)\right]\right.\nonumber\\
&&\times X_i(n)+\left.TQ(n)+P(1-Q(n))\right\}.
\end{eqnarray}

To clarify the function of fence-sitters, we consider a payoff memory to amplify their function to an observable level.
The payoff memory is widely adopted in a point system. The point system is composed by aggregated scores of individuals. The scores are aggregated for a number of rounds of games in a season.
Aggregating the payoff gaining from the starting round $n_0$ to round $n_0+\lambda-1$ ($\lambda>1$), one can derive three classes of averaged payoffs:
pure cooperators' average payoff ($\alpha$), pure defectors' average payoff ($\beta$) and fence-sitters' average payoff ($\gamma$).
\begin{equation}
\alpha(n)=\langle{k}\rangle\left[S\cdot \lambda-(R-S)\sum_{n=n_0}^{\lambda+n_0-1}Q(n)\right],
\end{equation}
\begin{equation}
\beta(n)=\langle{k}\rangle\sum_{n=n_0}^{\lambda+n_0-1}\left[(T-P)Q(n)+P)\right],
\end{equation}
and
\begin{eqnarray}
\gamma(n)=\langle{k}\rangle\left\{\sum_{n=n_0~and~X_i(n)=1}^{\lambda+n_0-1}\left[(R-T+P-S)Q(n)+(S-P)\right]+\sum_{n=n_0}^{\lambda+n_0-1}\left[TQ(n)+P(1-Q(n))\right]\right\}.
\end{eqnarray}

%\section{Payoff Memory and Cooperation}
For the PD, when the system reaches a relatively stable status for a certain updating rule, these three variables can be rewritten as
\begin{equation}
\alpha_{\infty}=\lambda\langle{k}\rangle \left[(R-S)Q(\infty)+S\right],\label{PCP}
\end{equation}
\begin{equation}
\beta_{\infty}=\lambda\langle{k}\rangle \left[(T-P)Q(\infty)+P\right],\label{PDP}
\end{equation}
and
\begin{equation}
\gamma^{\omega}_{\infty}\!\!\!=\!\lambda\langle{k}\rangle\!\!\left(\frac{S-P+(R-T+P-S){Q(\infty)}}{\omega}+(T-P)Q(\infty)+P\right),\label{FSP}
\end{equation}
where $\infty$ denotes $n\rightarrow\infty$ and $\omega\in\left[1,+\infty\right)$ denotes the average shifting period of cooperation. For example, if $\omega=3$, the fence-sitter plays cooperator once in three rounds on average. For the PD, $\alpha_{\infty}\leq \beta_{\infty}$. Before the extinction of pure cooperators in the population, pure defectors can not die out. Similarly, for $\gamma_{\infty}>\alpha_{\infty}$, if pure cooperators still exist in the population, fence-sitters can survive as well. In an evolutionary game, the number of fence-sitters is determined by the updating rules of strategies.
We define $F_{\infty}^\textrm{PC}$ and $F_{\infty}^{S(\omega)}$ as the frequency of the pure cooperators and fence-sitters with period $\omega$ in the relatively stable status respectively. One can derive the final frequency of cooperators
\begin{equation}
Q(\infty)=F_{\infty}^\textrm{PC}+\sum_{\omega}\frac{F_{\infty}^{S(\omega)}}{\omega}.\label{Qinf}
\end{equation}
The memory promotes the advantage of payoff for the fence-sitters. The higher payoff of the fence-sitters brings them a larger living space. The existence of more fence-sitters enhances the frequency of cooperators. Hence, the payoff memory indirectly promotes the level of cooperation.

\section{Simulations}
To test our theoretical prediction, we run the PD extensively on random graphs with a updating rule proposed by Santos and Pacheco~\cite{PRL95098104}. Santos's updating rule simulates a local random optimized process of strategy. In this process, an individual $i$ chooses a randomly picked neighbor $j$ as its reference. At the $n$th round, if $j$'s payoff is higher than that of $i$, $i$ will play $j$'s strategy in the next round with a probability directly proportional to the difference of their payoffs $\sum_{t=r\times \lambda+1}^n\left(G_j(t)-G_i(t)\right)$ and inversely proportional to $Max\{k_i,k_j\}\cdot T$. The parameter $r=\lfloor\frac{n}{\lambda}\rfloor-1$ for $\frac{n}{\lambda}\in\mathbb{N}$, while $r=\lfloor\frac{n}{\lambda}\rfloor$ for $\frac{n}{\lambda}\notin\mathbb{N}$, so that, one can find $t$ in the time region $\left[r\times \lambda+1,(r+1)\times \lambda\right]$.

\begin{figure}
\scalebox{0.5}[0.4]{\includegraphics[trim=0 50 0 0]{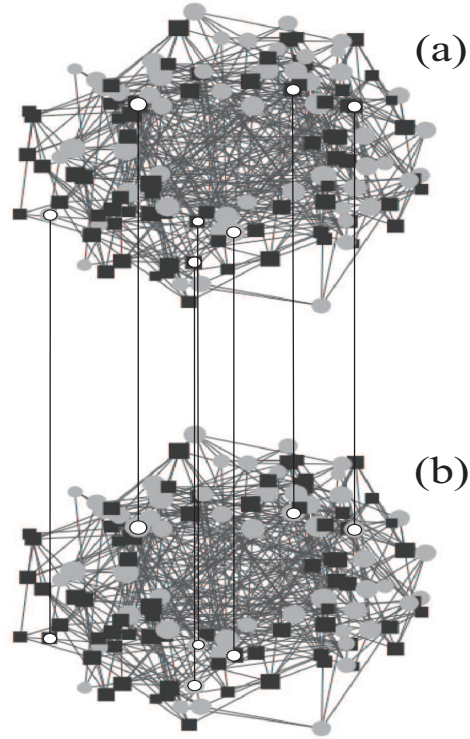}}
\caption{Illustrations of fence-sitters in networks. Light gray circles and dark gray squares denote cooperators and defectors respectively. Scales of individuals denote their degrees. The simulation results are obtained by the updating rules proposed by Santos and Pacheco~\cite{PRL95098104} for the PD with $T=1.2$, $P=S=0$ and $R=1$. We set $\lambda=1$ to show the fence-sitters in the original model.
(a) and (b) show the strategies of individuals from step $10~001$ to $10~002$. Hollow circles or squares connected by solid lines denote a part of the fence-sitters in question. One can observe their strategies in (a) are different from that in (b).
The random graphs are generated by randomly rewiring all of the links in the regular graphs, which are formed by $128$ identical individuals with average degree $6$.
}\label{F1}
\end{figure}

For different initial conditions, after a period of initial turbulence, network gaming always reaches a dynamical equilibrium, where the number of cooperators (or defectors) is stabilized at the particular value with minimum fluctuation. As shown in Fig.~\ref{F1}(a) and Fig.~\ref{F1}(b), fence-sitters are the individuals which change their color with time in the dynamical equilibrium. Comparing Fig.~\ref{F1}(a) with Fig.~\ref{F1}(b), one can observe that the colors of a couple of individuals shift from step $10~001$ to $10~002$, which are a part of the fence-sitters in question.

\begin{figure}
\scalebox{0.35}[0.33]{\includegraphics[trim=0 50 0 0]{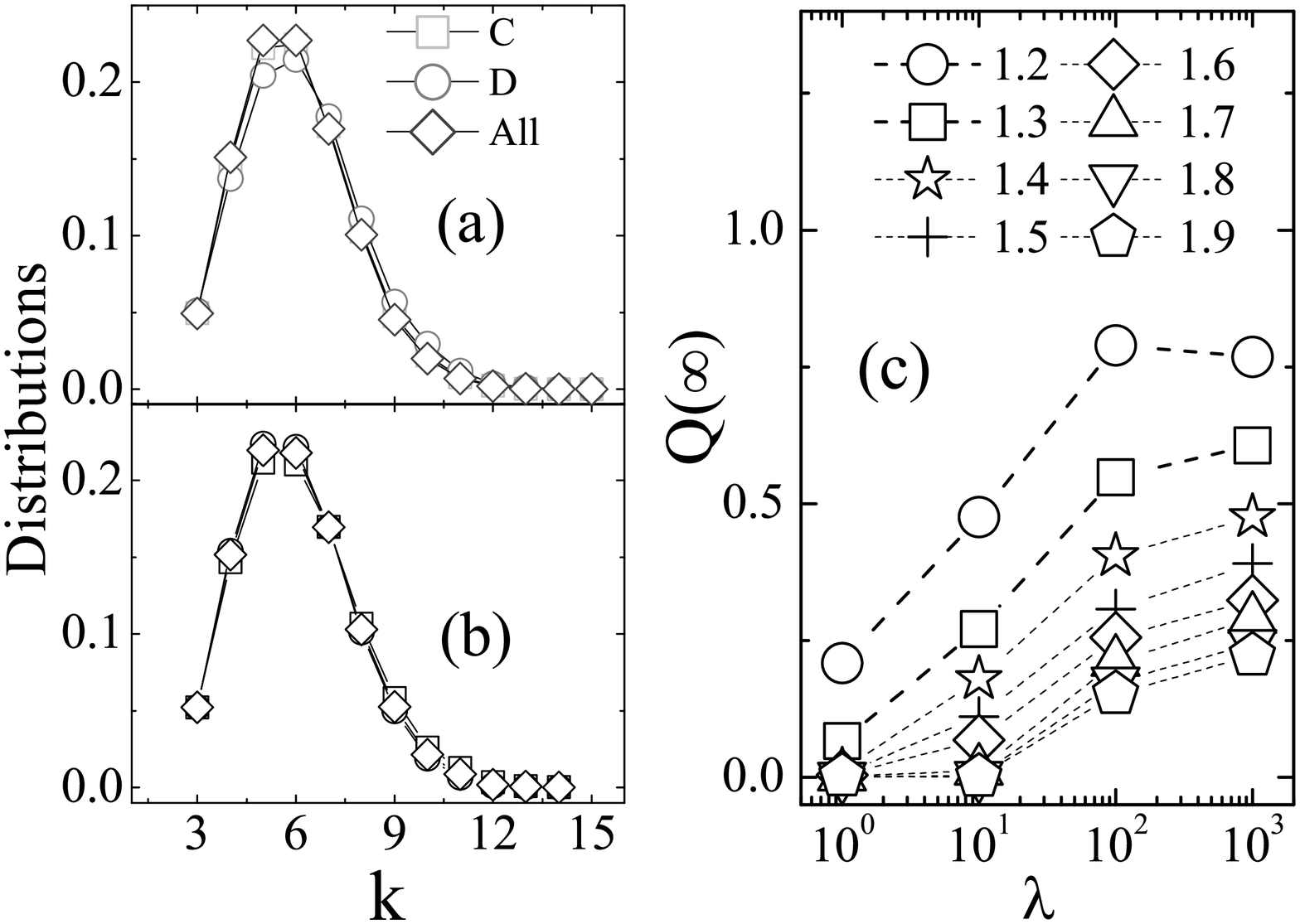}}
\caption{Degree distributions of cooperators (squares), defectors (circles) and all the individuals (diamonds). (a) and (b) show the simulation results for $T=1.2$ and $1.3$ respectively. The degree distributions show that cooperators are evenly distributed in the random graphs.
(c) shows the frequency of cooperators $Q(\infty)$ as a function of the payoff memory $\lambda$ for $T\in[1.2,1.9]$. In these simulations, the size of random graphs is set to $1024$.
}\label{F2}
\end{figure}

Figure~\ref{F2}(a) and Fig.~\ref{F2}(b) show the degree distributions of cooperators, defectors, and all the individuals, respectively. For both panels, one can find that the three distributions are overlapped basically, which is the condition of Eq.~(\ref{Approxiation}). Figure~\ref{F2}(c) shows the relation between the frequency of cooperators $Q(\infty)$ and the payoff memory $\lambda$. For $T\in[1.2,1.9]$, one can observe that the level of cooperation grows with $\lambda$ monotonously. For $T\leq1.1$, all the individuals are pure cooperators. For these cases, the simulation results are thus not shown in this figure.

\begin{figure}
\scalebox{0.36}[0.4]{\includegraphics[trim=0 30 0 0]{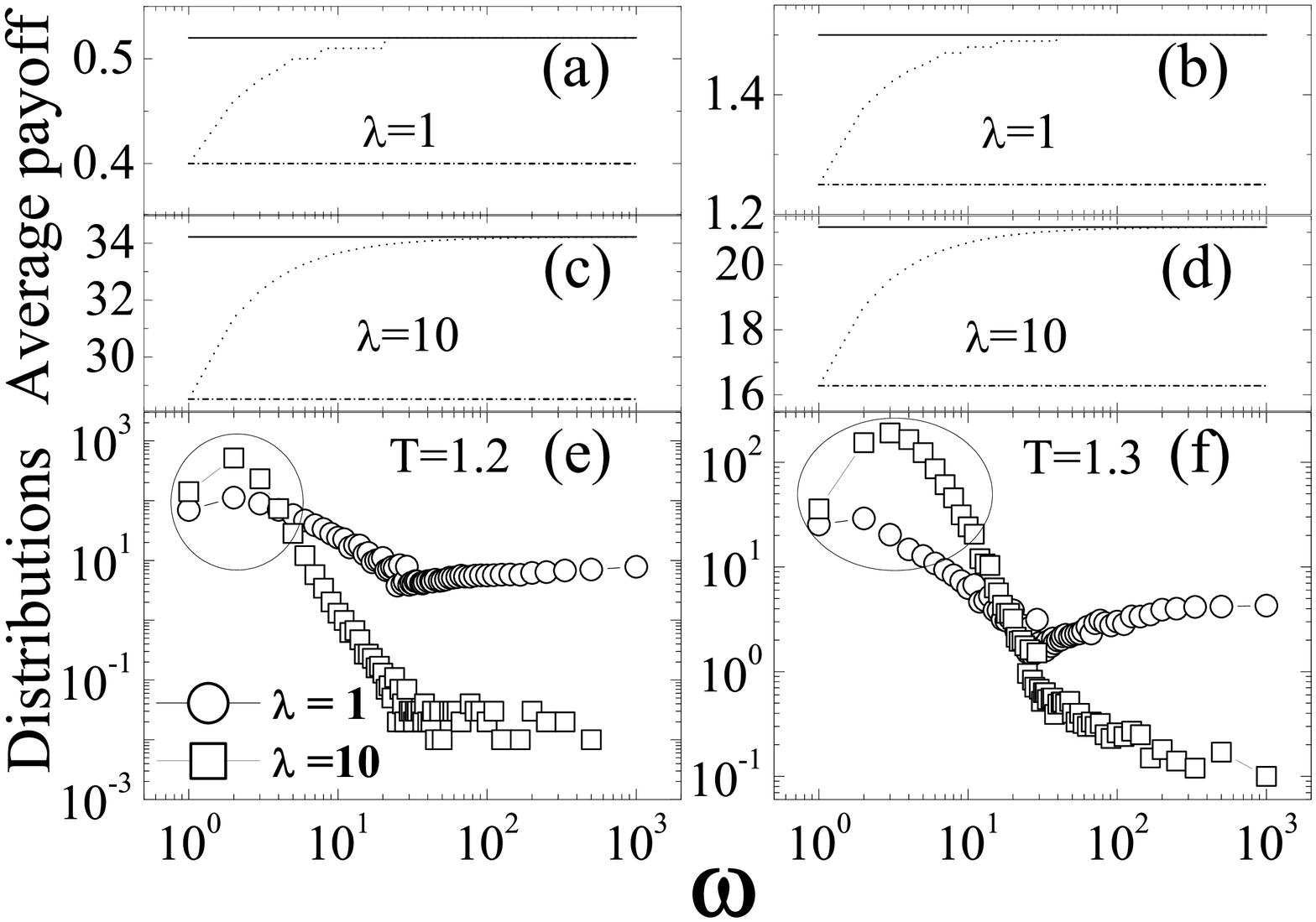}}
\caption{Average payoff of different classes of individuals and the distributions of average shifting period $\omega$ in random graphs.
(a), (b), (c), and (d) show the analytic results obtained by Eq.~(\ref{PCP}), Eq.~(\ref{PDP}), and Eq.~(\ref{FSP}). Solid lines, dotted lines, and dash-dot lines denote the payoff of the pure defectors, fence-sitters, and pure cooperators, respectively. (a) and (c) show the cases of $\lambda=1$ and $10$ for $T=1.2$, respectively. The analytic results are based on $Q(\infty)=0.21$ and $0.48$ for $\lambda=1$ and $10$, respectively. The values of $Q(\infty)$ are obtained by corresponding simulations in (e). Similarly, (b) and (d) show the cases of $\lambda=1$ and $10$ for $T=1.3$, respectively. The analytic results are based on $Q(\infty)=0.07$ and $0.27$ for $\lambda=1$ and $10$, respectively. The values of $Q(\infty)$ are obtained by corresponding simulations in (f). The random graphs are formed by $1024$ identical individuals with average degree $6$. We run ten simulations for each of the parameter values for the game on each of the ten networks. Thus each plot in the figure corresponds to $100$ simulations.
(e) and (f) show the distributions of $\omega$ for $T=1.2$ and $1.3$, respectively. (f) shares the same legend with (e). Comparing the data points of $\omega=1$ and $2$ in the ellipses, one can find the number of fence-sitters with $\omega\leq 10$ grows with $\lambda$ dramatically. This behavior originates from the fact that their payoffs are much higher than that of pure cooperators for $\lambda=10$. The differences of the payoffs are shown clearly in (a), (b), (c), and (d).
Note that (a), (b), (c), and (d) are semilog graphs. (e) and (f) are log-log graphs, in which the scales on distributions are different.
}\label{F3}
\end{figure}

Why does payoff memory enhance $Q(\infty)$? Based on Eq.~(\ref{Qinf}), one can find that there are only two possibilities. One is that more pure cooperators emerge; the other is that more fence-sitters emerge. To make clear the final reason, we measure the average payoff and distributions of individuals with different average shifting period $\omega$ in Fig.~\ref{F3}. Comparing Fig.~\ref{F3}(a) and Fig.~\ref{F3}(b) with Fig.~\ref{F3}(c) and Fig.~\ref{F3}(d), one can observe the dramatic increase of the differences between the dash-dot lines and dotted lines, which denote the pure cooperators and fence-sitters' payoff, respectively. The increase leads to a high probability for the fence-sitters' pure cooperative neighbors to copy the fence-sitters' strategy. Naturally, these pure cooperative neighbors turn out to be new fence-sitters in the later rounds. The number of fence-sitters grows with $\lambda$, but it has its upper bound. The upper bound is generated by their payoffs, which are always lower than that of the pure defectors. The pure defectors' payoffs are denoted by the solid lines in Fig.~\ref{F3}(a), Fig.~\ref{F3}(b), Fig.~\ref{F3}(c), and Fig.~\ref{F3}(d).
In Fig.~\ref{F3}(e) and Fig.~\ref{F3}(f), for $\lambda=1$, one can observe that the fence-sitters are more than the pure cooperators corresponding to $\omega=1$ in Fig.~\ref{F3} for both $T=1.2$ and $T=1.3$. For $\lambda=10$, more fence-sitters emerge in both panels. The increase of fence-sitters originates from the nonlinear increase of the difference between their average payoff and the pure cooperators' average payoff.

Notice that the plots are binned in this figure; a large number of fence-sitters with a short average shifting period ($\omega<1.5$) serve as the pure cooperators ($\omega=1$). Actually, for the games with a long payoff memory and a large $T$, the cooperators in the dynamical equilibrium are mainly composed of the cooperating fence-sitters, that is, the second term in Eq.~(\ref{Qinf}). Figure~\ref{F3} indicates that the fitness of fence-sitters is not less than the pure cooperators for all the parameters investigated here. When $\lambda>1$, the advantage of payoff highly promotes the frequency of fence-sitters. Based on Eq.~(\ref{Qinf}), the promotion on the frequency of fence-sitters leads to a remarkable increase on $Q(\infty)$.

The example of random graph suggests that the payoff memory can highly promote the level of cooperation in a type of network. It does not mean that the influence only exists in this type of network. For other networks, such as Watts and Strogatz small-world networks~\cite{NATURE393440}, Barab\'{a}si and Albert's scale-free networks~\cite{SCIENCE286509}, and regular networks, the influences are still considerably apparent. In addition, we also observe a series of similar results for the updating rule proposed by Nowak and May~\cite{NATURE359826}. All these observations indicate that fence-sitters can promote the level of cooperation in evolutionary games in complex networks. Note that our vectorial formalization can also be applied to other two-strategies games, such as the hawk-dove game (also known as the snow-drift or chicken game)~\cite{ETG,GTE,NATURE428643,EL8748}, Hisashi Ohtsuki's model~\cite{NATURE441502} and so forth.

\section{Conclusion}%{Conclusion}
In summary, we have investigated the role of fence-sitters in evolutionary games in complex networks. The fence-sitters are the individuals who change their strategies at least once in the strategy evolutionary process. We introduce the first vectorial formulation to derive three classes of individuals' payoff analytically. The three classes are pure cooperators, pure defectors and fence-sitters. To clarify the function of the fence-sitters, we define a control parameter, payoff memory, as the number of rounds that the individuals' payoffs are aggregated. We observe that the fence-sitters' effects and the level of cooperation grow with the payoff memory apparently.
%Previous studies
Previous studies concentrated on the influences of topological structures on the level of cooperation in the complex networks. In these numerical studies, the fence-sitters were observed in all the systems.
%Our contributions

In this paper, we have analytically demonstrated that the existence of fence-sitters can promote the level of cooperation. We have introduced a vectorial formulation of individuals' payoffs, which can be applied to all the two-strategies games. We have also proposed a mean-field approach to derive these payoffs analytically. For the prisoner's dilemma, we found the average payoff of the individuals with a shifting strategy is not less than the pure cooperators' payoff. Because of their advantage on payoff, the fence-sitters are more robust than the pure cooperators to the invasion of pure defectors. Although the fence-sitters' payoff is lower than that of the pure defectors, pure cooperators tend to become fence-sitters first. Before the extinction of cooperators, the majority of cooperators are composed of a part of the fence-sitters. Thus the existence of fence-sitters promotes the fitness of cooperators in an indirect way.
%Meaning of our work
In complex networks, our observations indicate that cooperation is protected by the individuals with a shifting strategy.  Our results may provide a better understanding of the composition of cooperators in a circumstance where defectors always obtain a higher payoff.

%{Acknowledgment}
Y. Z. and M. A. A.-A. and C. B. are supported by the region Haute Normandie, France, and the ERDF RISC. S. Z. is supported by the UK Royal Academy of Engineering and the Engineering and Physical Sciences Research Council (EPSRC) under Grant No. 10216/70.

%%%%%%%%%%%%%%%%%%%%%%%%%%%%%%%%%%%%%%%%%%%%%%%%%%%%%%%%%%
% Figure 1
%%%%%%%%%%%%%%%%%%%%%%%%%%%%%%%%%%%%%%%%%%%%%%%%%%%%%%%%%%
%%%%%%%%%%%%%%%%%%%%%%%%%%%%%%%%%%%%%%%%%%%%%%%%%%%%%%%%%%

%%%%%%%%%%%%%%%%%%%%%%%%%%%%%%%%%%%%%%%%%%%%%%%%%%%%%%%%%%
% Figure 2
%%%%%%%%%%%%%%%%%%%%%%%%%%%%%%%%%%%%%%%%%%%%%%%%%%%%%%%%%%
%%%%%%%%%%%%%%%%%%%%%%%%%%%%%%%%%%%%%%%%%%%%%%%%%%%%%%%%%%

%%%%%%%%%%%%%%%%%%%%%%%%%%%%%%%%%%%%%%%%%%%%%%%%%%%%%%%%%%
% Figure 3
%%%%%%%%%%%%%%%%%%%%%%%%%%%%%%%%%%%%%%%%%%%%%%%%%%%%%%%%%%
%%%%%%%%%%%%%%%%%%%%%%%%%%%%%%%%%%%%%%%%%%%%%%%%%%%%%%%%%%

\end{document}